\documentstyle[12pt]{article}
\newcommand{\wm}{\widehat{\mu}}
\newcommand{\wn}{\widehat{\nu}}
\newcommand{\wi}{\widehat{i}}
\newcommand{\wj}{\widehat{j}}
\newcommand{\mn}{\mu\nu}
\begin{document}
\title{The Calculation of the Perturbative Expansion of Wilson Operators
on Lattice }

\author{{Da Qing Liu$^2$ and Ji Min Wu$^{1,~2}$}\\
        {\small $^1$CCAST(World Lab. ), P. O. Pox 8730, Beijing 100080,
                China}\\
        {\small $^2$Institute of High Energy Physics, Chinese Academy
                of
 Sciences,}\\
        {\small P. O. Box 918-4, Beijing, 100039, P. R. China}
       }
\maketitle

\begin{center}
\begin{minipage}{5.5in}
\vskip 0.8in
{\bf Abstract}\\
We introduce an approach to expand gauge-invariant Wilson operators
on lattice. This approach is based on non-abelian Stokes theorem and
overcomes some shortage of some former methods. It is also suitable 
for expanding any Wilson operator on lattice.
\end{minipage}
\end{center}
\vskip 1in
\indent

\newpage
\section {Introduction}
Expanding gauge-invariant Wilson operators, or, Wilson loops, according to
lattice spacing $a$ is a very important work when we do theoretical
analysis in lattice QCD. It is a necessary step in many cases, such as
the discussion of improved action$^{\cite{s1}}$, the exploration
of improved cooling$^{\cite{s2}}$ or the investigate for renormalization
procedure$^{\cite{s3}}$ when study instanton phenomenon, etc.
Furthermore, in order to calculate masses of glueball states, we developed
a new opinion$^{\cite{s4}}$, in which the perturbative
expansion of Wilson loops is also a needed progress.

Due to the complicated of path ordering operators, it is cockamamie
and difficult to expand the Wilson loops according to normal
method if the number of links of Wilson loop is large or it is
needed to expand Wilson loop up to higher precision. There is a prevalent
approach introduced by Luscher and Weisz in ref.\cite{s5}. But
this approach has some shortcoming, for instance, one
is likely to face some difficult when he tries to expand
relatively complicated spatial loops.

We introduce a practical easy approach which is based on the
non-abelian Stokes theorem here. This approach has overcome the
shortcoming in ref. \cite{s5} and is
therefore suitable for any Wilson loops on lattice. It is also suitable for any of above 
cases in ref. \cite{s1,s2,s3,s4}. 
To do this, we
will give a short comment on Luscher-Weisz method in the following
section and some mathematical preparation in section 3. We
exemplify our method in section 4.

Since the case in ref. \cite{s4} is the most difficult and
universal one, we should pay our most attention on this case.

\section {The Comment on Luscher-Weisz Method}
Let us observe an arbitrary gauge-invariant Wilson loop on the
cubic lattice. In order to get the perturbative expansion of the
loop up to $O(a^6)$ ( To simplify, this paper only discusses the
expansion up to $O(a^6)$), Luscher and Weisz adopt the following
method: they fixed a gauge at first, then try to rewrite it in
gauge-invariant form.

Assume that one needs to expand the plaquette operator:
\begin{equation}\label{e1}
pl=\sum\limits_n\sum\limits_{{\mu}{\nu}}pl_{{\mu}{\nu}}(n)=\sum\limits_n
\sum\limits_{{\mu},{\nu}}Tr\{U(n,{\mu})U(n+\wm ,{\nu})
U^{-1}(n+\wn ,{\mu})U^{-1}(n,{\nu})\},
\end{equation}
where $\wm$ is the unit vector along ${\mu}$ direction, $U(n,{\mu})$
is link variable defined as follows:
$$ U(n,{\mu})=T\exp a\int\limits_0^1 dtA_{\mu}(an+a\wm t), $$
in which $A_{\mu}$is gauge potential\footnote{
We use a notation here that $A_{\mu}$ is anti-hermitian, traceless
$N\times N$-matrix}. We point out here that the summation is different
following different problem. For instance, the summation of $n$ is
over all the 4-dimensional lattice sites and the summation of
directions( $\mu\nu$) is over all possible 4-dimensional space-time
directions when one discusses the improved action$^{\cite{s1}}$, while
the summation of $n$ is over spatial lattice sites in the same time slice
and that of directions is variant in 3-dimensional spatial directions
according to different problems in the case of ref. \cite{s4}.

One can always choose a proper local gauge on lattice to make the
gauge field satisfy
\begin{eqnarray} \label{e2}
 A_{\mu}(x)=0,& \mbox{for all }x, \nonumber \\
A_{\nu}(x)=0,& \mbox{for all~} x \mbox{ with } x_{\mu}=0.
\end{eqnarray}
$pl_{\mn}(0)$ then reduces to
\begin{equation}\label{e3}
pl_{\mn}(0)=TrT\exp a\int\limits_0^1 dtA_{\mu}(a\wm +a\wm t).
\end{equation}
At the same time, at $x=0$, one has
\begin{equation}\label{e4}
\partial_{\nu}^p\partial_{\mu }^{q+1}A_{\nu }=D_{\nu }^pD_{\mu }^qF_{\mn}.
\end{equation}
Insert Eq. (\ref{e4}) into Eq. (\ref{e3}) and expand Eq.
(\ref{e3}) according to lattice spacing $a$:
 \begin{eqnarray}
 pl_{\mn}(0)&=&N+{1\over
2}a^4Tr(F_{\mn}F_{\mn})+{1\over 2}a^5Tr
[F_{\mn}(D_{\mu}+D_{\nu})F_{\mn}]
\nonumber \\
~&+&{1\over 6}a^6Tr\{F_{\mn}^3+F_{\mn}(D_{\mu}^2+{3\over 2}
D_{\mu}D_{\nu}+D_{\nu}^2)F_{\mn} \nonumber \\
\label{e5}
&+&{3\over 4}(D_{\mu}+D_{\nu})F_{\mn}(D_{\mu}+D_{\nu})F_{\mn} \}+O(a^7),
\end{eqnarray}
where $F_{{\mn}}=\partial_{\mu}A_{\nu}-\partial_{\nu}A_{\mu}+[A_{\mu},A_{\nu}]$
and
$D_{\mu}F_{\nu \rho}=\partial_{\mu}F_{\nu \rho}+[A_{\mu},F_{\nu
\rho}]$. Due to periodic boundary condition, one can ignore all the
(covariant) differential terms in the last result( Section 4 shows
more details) and gets:
\begin{equation}\label{e6}
pl_{\mn}(0)={1\over 2}a^4TrF_{\mn}^2-{1\over
24}a^6Tr[F_{\mn}(D_{\mu}^2+D_{\nu}^2)F_{\mn}]+
{1\over 6}a^6TrF_{\mn}^3+O(a^7).
\end{equation}
Now let us observe Eq. (\ref{e4}). In this case, because
\begin{equation}\label{e7}
D_{\mu}D_{\nu}-D_{\nu}D_{\mu}=[F_{{\mn}},\cdot],
\end{equation}
the corresponding in Eq. (\ref{e4}) is unambiguous. But, when one
tries to expand some complicated spatial Wilson operators, l.h.s. of Eq.
(\ref{e4}) is possible as the form
$\partial_{\mu}\partial_{\nu}A_{\rho}(\mu\neq\nu\neq\rho)$.
There is ambiguity that such
terms may correspond both to $D_{\mu}F_{\nu\rho}$ and
to $D_{\nu}F_{\mu\rho}$, which means one can not determine the
expansion unambiguously, especially when the summation of
directions is not over all possible directions but some definite
ones$^{\cite{s4}}$.

In the case of ref. \cite{s1,s5} that
one should sum over all possible 4-dimensional directions, Luscher and
Weisz eliminate the ambiguity through the presumption that the
expansion of any Wilson loop after summation over all possible
directions and lattice sites is invariant under any Lorentz
transformation( In 4-dimensional
Euclidean space, it is $SO(4)$ group), parity
and charge conjugacy transformation. This presumption will some
time complicate our expansion, furthermore, it has not been proved
strictly. We show a debate here. Suppose we accepts this
presumption and restricts ourselves into 3-dimensional space, the
subspace of the 4-dimensional Euclidean space and time. One
natural deduction is that the operator, which has been summed over
all space lattice sites in the same time slice and all possible
3-dimensional directions,
is invariant under $SO(3)^{PC}$ group since the $SO(3)$ group is the
subgroup of the $SO(4)$ group. As such operator belongs to
representation $A_1^{++}$, the trivial representation of the cubic
group on lattice, one may regard that the $J^{PC}$( in continuum limit)
of the operator belonging to $A_1^{++}$ is $0^{++}$. In other words, the
subduced
representation of the representation $0^{++}$ of the $SO(3)^{PC}$
group can be reduced into $A_1^{++}$( subduced representation: the
representation which is obtained by trivially embedding the
cubic group $O^{PC}$ into the group $SO(3)^{PC}$). But, as we
know, there also exist other representations in the group
$SO(3)^{PC}$, the subduced representation of which can be reduced
into $A_1^{++}$, for instance, $4^{++}$ $^{\cite{s6,s7}}$, which
means that the $J^{PC}$( in the continuum limit) of the operator
belonging to $A_1^{++}$ is
 not always $0^{++}$. So, the validity of the presumption is
 doubtful. In fact, a discussion in ref. \cite{s13} does not support this 
presumption too. To say the least,
even this presumption is right, one
 can hardly apply it to expand the operator, for instance, the
 $PC$ of which is not $++$.

To solve the problem, we developed a new approach which can
determine the perturbative expansion unambiguously. The next section is
the proper mathematical preparation.

\section {Mathematical Preliminary}
For operators, or loops, the traces over color
space are not performed in this section, so, these operators are gauge
covariant under gauge transformation.

Now we begin our preparation with some simple examples.

Observe the operator $\tilde{W}(O,Z)$ shown in Fig. 1:
\begin{equation}\label{e8}
  \tilde{W}(O,Z)=U_R(Z,O;C_X)W(Z)U_R(O,Z;\bar{C}_X),
\end{equation}
where $W$ is a definite gauge covariant operator which is from point $Z$ and back to
$Z$, and $U_R(Z,O,C_X)=P\exp\int_0^xdtA_i(\widehat{i}t)$ is
connector. Some properties of connectors are list as
follows$^{\cite{s12}}$:
\begin{description}
  \item[a)]
\begin{equation}
 U_R(X_3,X_1;C_1+C_2)=U_R(X_2,X_1;C_1)\cdot U_R(X_3,X_2;C_2),
\end{equation}
where $C_1$ is a curve going from point $X_1$ to point $X_2$ and $C_2$
is a curve going from point $X_2$ to point $X_3$.
  \item[b)]
\begin{eqnarray}
  U_R(X,Y;\bar{C})U_R(Y,X;C)=1, \nonumber \\
  U_R^{\dag}(Y,X,C)=U_R(X,Y;\bar{C}),
\end{eqnarray}
$C$ is a curve from $X$ to $Y$ and $\bar{C}$ the same curve but
oriented in inverse direction, running from $Y$ to $X$.
\end{description}
From the opinion of differential geometry, $\tilde{W}$ is the
shifted version of operator $W$ from point $Z$ to point $O$.

Under local gauge transform $V$ we have:
\begin{eqnarray}
\tilde{W}(O,Z) &\rightarrow& V(O )\tilde{W}(O,Z)V^{-1}(O),
\nonumber \\
\label{e9}
W(Z) &\rightarrow& V(Z)W(Z)V^{-1}(Z).
\end{eqnarray}
Let $W(Z)=W+x\partial_iW+\cdots+{x^n \over n!}\partial_i^{n}W+\cdots$,
where $W\equiv W(O)$ is the same operator as $W(Z)$ but it
is from $O$ and back to $O$.

We can prove that
\begin{equation}\label{e10}
\tilde{W}(O,Z)=W+xD_iW+\cdots+{x^n \over n!}D_i^nW+\cdots .
\end{equation}

Then we study the operator described in Fig. 2 ( We also denote it
by $\tilde{W}$):
\begin{equation}\label{e11}
  \tilde{W}(O,Z)=U_R(Z,O;C_Y+C_X)W(Z)U_R(O,Z;\bar{C}_X+\bar{C}_Y).
\end{equation}
We let $W(Z)=W+x\partial_iW+y\partial_iW+{1\over 2}(x^2\partial_i^2+
2xy\partial_i\partial_jW+y^2\partial_j^2W)+\cdots$, in which
$W\equiv W(O)$. Using Eq. (\ref{e10}) and the properties of
connector we get
\begin{equation}\label{e12}
\tilde{W}(O,Z)=W+xD_iW+yD_jW+{1\over
2}(x^2D_i^2+2xyD_jD_i+y^2D_j^2)W+\cdots.
\end{equation}
It should be pointed out here that, in general, the
order of $D_jD_i$ can not be reversed in above equation.

Other more complicated case can be calculated as similar as above.

We write out non-abelian Stokes theorem as( We let $t_0\equiv
a$)$^{\cite{s8}}$
\begin{eqnarray}
U_{\mu
\nu}(x)&=&U_{\mu}(x)U_{\nu}(x+\wm)U_{\mu}^{\dag} (x+\wn)U_{\nu}^{\dag}
(x) \nonumber \\
&=&P\exp
\int_0^ads\int_0^adt\tilde{F}_{\mn}(x+s\wm+\wn)
\nonumber \\
&=&1+\sum\limits_{m=1}^{\infty}\prod\limits_{i=1}^m\int_0^a ds_i\int_0^{t_{i-1}}dt_i
\tilde{F}_{\mn}(x+s_n\wm+t_n\wn)\cdots
\tilde{F}_{\mn}(x+s_1\wm+t_1\wm).
\nonumber \\
\label{e13}
 &&
\end{eqnarray}
Here $\tilde{F}_{\mn}(x+s\wm+t\wn)$ is the shifted version of
$F_{\mn}(x+s\wm+t\wn)$ from point $x+s\wm+t\wn$ to point $x$, or:
$$
V(s,t)=P\exp\int_0^tdt'A_{\nu}(x+t'\wn)\cdot
P\exp\int_0^sds'A_{\mu}(x+t\wn+s'\wm),
$$
\begin{equation}\label{e14}
\tilde{F}_{\mn}(x+s_n\wm+t_n\wn)=V(s,t)F_{\mn}(x+s\wm+t\wn)V^{\dag}
(s,t).
\end{equation}
Now using Eq. (\ref{e12}), (\ref{e13}), (\ref{e14}) we can get the
expansion of the operator from $O$ back to $O$ which is shown
in fig. 3 (a):
\begin{eqnarray}\label{e15}
U^{(1)}_{ij
}(n)&=&U_i(n)U_j(n+\wi)U_i^{\dag} (n+\wj)U_j^{\dag}
(n) \nonumber\\
 & =&1+a^2F_{ij}+{a^3\over 2}(D_i+D_j)F_{ij}+
  {a^4\over 6}(D_i^2+D_j^2+{3\over 2}D_jD_i)F_{ij}
\nonumber \\
  &&+P_{ij}+{a^6\over 6}F_{ij}^3+O(a^7),
\end{eqnarray}
where $P_{ij}={a^4\over 2}F_{ij}^2+{a^6\over
24}F_{ij}(D_i^2+D_j^2)F_{ij}$,~$F_{ij}=F_{ij}(n)$.
Because our precision in last result is $O(a^6)$, we have ignored
all the full (covariant) differential terms the ranks of which are
large than or equal to $O(a^5)$( More details are shown in section 4), for
instance the term  ${a^5\over 2}F_{ij}(D_i+D_j)F_{ij}$.

Using Eq. (\ref{e15}) and the properties of connector we get the
perturbative expansion of the operator in Fig. 3(b):
\begin{eqnarray}\label{e16}
U^{(2)}_{ij}(n)&=&U_j(n)U_i^{\dag}(n-\wi+\wj)U_j^{\dag}(n-\wi)U_i(n-\wi)
\nonumber\\
 & =&1+a^2F_{ij}-{a^3\over 2}(D_i-D_j)F_{ij}+
  {a^4\over 6}(D_i^2+D_j^2-{3\over 2}D_jD_i)F_{ij}
\nonumber \\
  &&+P_{ij}+{a^6\over 6}F_{ij}^3+O(a^7).
\end{eqnarray}
With the same reason we get the perturbative  expansion of the operator in Fig.
3(c):
\begin{eqnarray}\label{e17}
U^{(3)}_{ij}(n)&=&U_i^{\dag}(n-\wi)U_j^{\dag}(n-\wi-\wj)U_i(n-\wi-\wj)U_j(n-\wj)
\nonumber\\
 & =&1+a^2F_{ij}-{a^3\over 2}(D_i+D_j)F_{ij}+
  {a^4\over 6}(D_i^2+D_j^2+{3\over 2}D_jD_i)F_{ij}
 \nonumber \\
   &&+P_{ij}+{a^6\over 6}F_{ij}^3+O(a^7).
 \end{eqnarray}
And the expansion of the operator in Fig. 3(d):
\begin{eqnarray}\label{e18}
U^{(3)}_{ij}(n)&=&U_j^{\dag}(n-\wj)U_i(n-\wj)U_j(n+\wi-\wj)U_i^{\dag}(n)
\nonumber\\
 & =&1+a^2F_{ij}+{a^3\over 2}(D_i-D_j)F_{ij}+
  {a^4\over 6}(D_i^2+D_j^2-{3\over 2}D_jD_i)F_{ij}
\nonumber \\
  &&+P_{ij}+{a^6\over 6}F_{ij}^3+O(a^7).
\end{eqnarray}
One can prove that the expansion of the charge conjugacy 
parity of above operator $U^{(i)}$ is equivalent to replace $F_{ij}$ by
$-F_{ij}$ in Eq. (\ref{e15}) - (\ref{e18}).

Now we can expand arbitrary Wilson loop on lattice using above
equations. It is shown in the following section.

\section {The Expansion of Arbitrary Wilson Loop on Lattice}
We now give a general properties of the perturbative expansion of
Wilson loops on lattice before we perform the expansion.
\begin{description}
  \item[a]~~~ Because
  differential operator can be presented by combination of difference
  operators with various ranks, the final contributions of full differential
  terms are zero due to the periodic boundary condition and we shall ignore
  such full differential terms in the expansion. In this reason, one can
  also expand Wilson loops on any lattice site with the same result.

  In this reason, one can ignore full (covariant) terms with rank
  large than or equal to $a^5$ in section 2 and 3 in our case.
  \item[b]~~
  On lattice an arbitrary Wilson loop can be expanded as
  \begin{equation}\label{e25}
  O(n)=\sum\limits_{k=0}^{\infty}a^kO_k(n),
  \end{equation}
  where $O_k(n)$, with mass-dimension $k$, is gauge-invariant
  polynomial of $A_{\mu}$ and $F_{\mn}$ and their derivatives.
  In above equation even $k$ corresponds to parity $P=+$ while odd
  $k$ correspond to parity $P=-$.
  \item [c]~~
  As we know, the $C$ parity of the real part of the operator is positive
  and that of the imaginary part is negative. Since the transformation of
  covariant differential $D_i$($D_i \cdot=\partial_i \cdot-i[A_i,\cdot]$)
  can be regard to be $1^{-+}$ under $SO(3)^{PC}$ group in this case, we
  can regard $D_i$ as a real operator in the expansion.
\end{description}
One can simplify the expansion using three above properties. For
instance, from above we know that $P_{ij}$ corresponds to $PC=++$
and it is invariant under translation on lattice if we only expand operators up
to precision $O(a^6)$.

Utilizing the perturbative expansion of $U^{(i)}(i=1,~2,~3,~4)$
in (\ref{e15})~-~(\ref{e18}) one can expand an arbitrary Wilson loop
on lattice. we exemplify it here.

The expansion of operators with $PC=++$ is the most complicated and
we should only focus on it. So, when we say an gauge-invariant operator
$L$ here we actually means ${1\over
4}(1+\widehat{C}+\widehat{P}+\widehat{C}\widehat{P})L$, where $\widehat{P}$
and $\widehat{C}$ are parity operator and charge conjugacy parity operator
which acts on operator $L$.

Let us observe the operator described in Fig. 4 and only care terms
with $PC=++$, so that we may only keep real terms with even $k$ in Eq.
(\ref{e25}).
\begin{equation}
L=Tr(1\cdot2\cdot3\cdot\bar{4}\cdot\bar{5}\cdot\bar{6}\cdot\bar{7}\cdot8),
\end{equation}
where, for simplification, we denote, for instance, the link variable $U(1)$ by $1$ and
$U^{\dag}(1)$ by $\bar{1}$, etc.

We rewrite $L$ as
\begin{eqnarray}
L&=&Tr[1\cdot2\cdot3\cdot\bar{4}\cdot\bar{9}\cdot\bar{1}
\cdot1\cdot9\cdot\bar{5}\cdot\bar{10}
\cdot10\cdot\bar{6}\cdot\bar{7}\cdot8]
\nonumber \\ \label{e19}
&=&Tr[pl_1\cdot pl_2\cdot pl3],
\end{eqnarray}
where $pl_1=1\cdot2\cdot3\cdot\bar{4}\cdot\bar{9}\cdot\bar{1}$,
$pl_2=1\cdot9\cdot\bar{5}\cdot\bar{10}$ and
$pl_3=10\cdot\bar{6}\cdot\bar{7}\cdot8$.
 Each of the $pl$ is gauge covariant respect to point $O$ under
local gauge transformation.

Using equations in section 3, we get
\begin{eqnarray}\label{e20}
  pl_1&=&1+a^2F_{ik}+{a^3\over 2}(D_i+D_k+2D_j)F_{ik}
  \nonumber \\
  &&+{a^4\over 6}(D_i^2+D_k^2+3D_j^2+{3\over
  2}D_{ik}+3D_{ji}+3D_{jk})F_{ik}+P_{ik}+O(a^7),
\end{eqnarray}
where $D_{ik}=D_iD_k$, etc.
\begin{equation} \label{e21}
  pl_2=1+a^2F_{jk}+{a^3\over 2}(D_j+D_k)F_{jk}+{a^4\over
  6}(D_j^2+D_k^2+{3\over 2}D_{jk})F_{jk}+O(a^7).
\end{equation}
and
\begin{equation}\label{e22}
  pl_3=1+a^2F_{ik}-{a^3\over 2}(D_i-D_k)F_{ik}+{a^4\over
  6}(D_i^2+D_k^2-{3\over 2}D_{ik})F_{ik}+O(a^7).
\end{equation}
So, up to $O(a^6)$, we have:
\begin{eqnarray}\label{e23}
  L&=&1+2a^4Tr(F_{ik}F_{ik})+{a^4\over
  2}Tr(F_{jk}F_{jk})+2a^4Tr(F_{ik}F_{jk})
  \nonumber \\
  &&+{a^6\over 6}Tr[F_{ik}(4D_i^2+D_k^2+3D_j^2+6D_{ij})F_{ik}]
  +{a^6\over 24}Tr[F_{jk}(D_j^2+D_k^2)F_{jk}]
  \nonumber \\
  &&+{a^6\over
  6}Tr[F_{ik}(2D_i^2+D_j^2+D_k^2+3D_{ij})F_{jk}]+O(a^7).
\end{eqnarray}
Here we have ignored terms with $PC\neq++$ and full covariant
differential terms.
Of course, we can do as similar for terms with $PC\neq ++$ in
the expansion.

\section{Conclusion}
We show an approach in the paper for the expansion of the Wilson
loops on lattice. The approach presented here can determine the
perturbative expansion of Wilson
loop on lattice unambiguously according to $P$ and $C$
 and is suitable for any Wilson loop
on lattice despite how tanglesome the Wilson loop is.

It is also an appropriate method to discuss the
improved action or improved cooling on lattice, etc.
In fact, in some of the former papers which discuss improved action,
there maybe exist some terms in higher precision
which are not
invariant under the rotation in $SO(4)$ group as
the continuum theory requires, and therefore should bring up some artifact
on lattice. Using this approach may help us to eliminate those unwanted
terms if we discuss the improved action or other interesting problem
up to higher precision.

\newpage
\section* {Figure captions}
\begin{itemize}
  \item
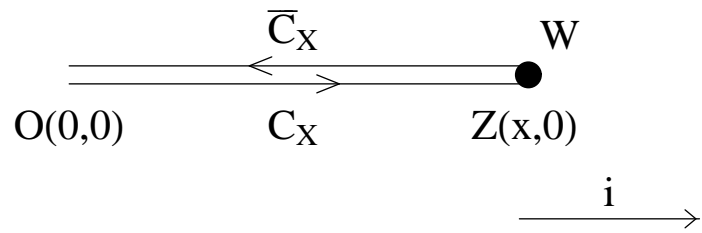
  {\bf Figure 1}~~~ The operator
   $U_R(Z,O;C_X)W(Z)U_R(O,Z;\bar{C}_X)$, where $U_R$ is the connector
   described in the paper.

  \item
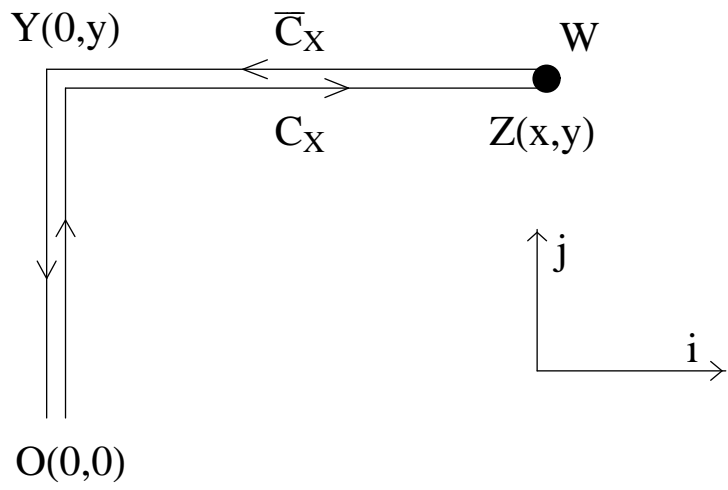
   {\bf Figure 2}~~~ The operator
   $U_R(Z,O;C_Y+C_X)W(Z)U_R(O,Z;\bar{C}_X+\bar{C}_Y)$, where the
   connector $U_R(Z,O;C_Y+C_X)=U_R(Y,O;C_Y)\cdot U_R(Z,Y;C_X)$.

  \item
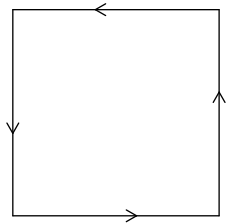
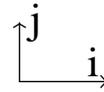
  {\bf Figure 3}~~~ The operators $U^{(i)}$ which are from point $O$
  and back to point $O$. They are of the same
    shape with the different starting and ending point.

\item
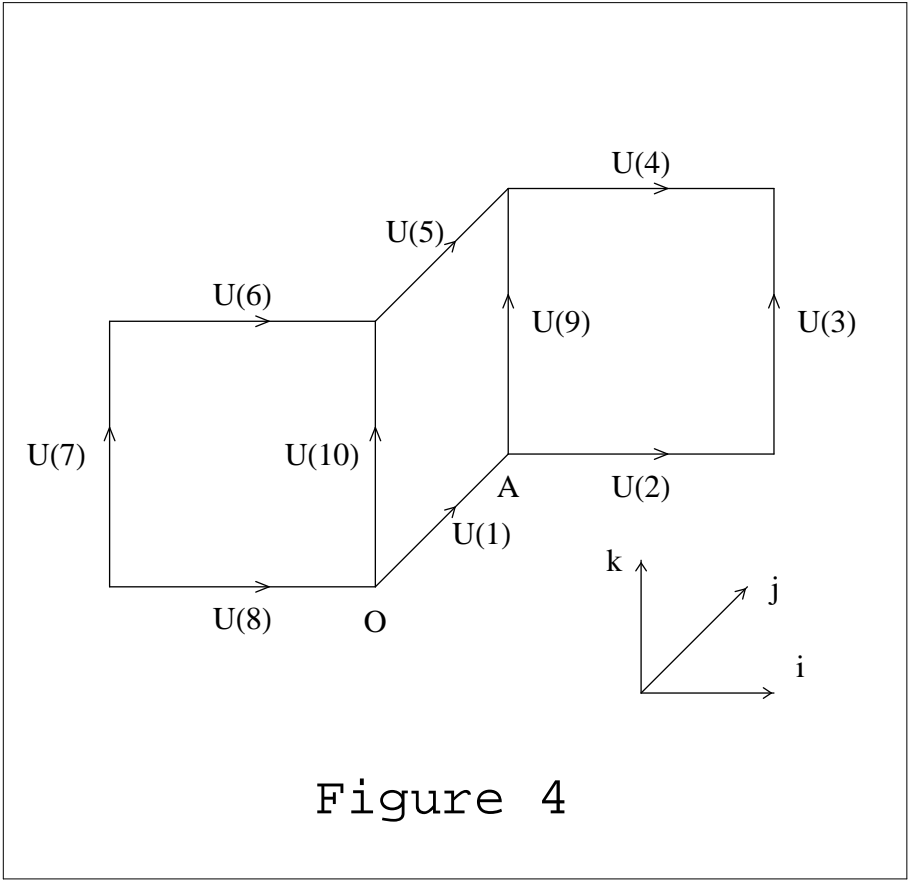
{\bf Figure 4}~~~ The operator
 $L=Tr(1\cdot2\cdot3\cdot\bar{4}\cdot\bar{5}\cdot\bar{6}\cdot\bar{7}\cdot8)$.
 Here we denote the link variable $U(i)(i=1,2,\cdots,10)$
  by $i$ and $U^{\dag}(i)$ by $\bar{i}$.

\end{itemize}

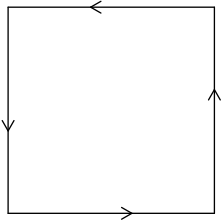
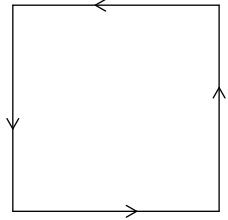
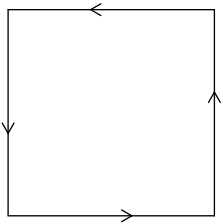
\end{document}